 \documentclass[preprint]{aastex}

\slugcomment{NAOJ-Th-Ap 2000,No.27}

\shorttitle{Thermal Instability}
\shortauthors{Fujita}

\begin{document}

\title{The Maximum Effect of Thermal Instability on Galactic Outflows}

\author{Yutaka Fujita}
\affil{National
Astronomical Observatory, Mitaka, Tokyo 181-8588, Japan}
\email{yfujita@th.nao.ac.jp}

\begin{abstract}
 We have investigated steady, radial gas outflows (or winds) from
 galaxies and the development of thermal instability in the hot gas. In
 order to see the maximum influence of the instability on the global
 structure of the galactic outflows, we study inhomogeneous comoving
 flows and the non-linear fate of the fluctuations in the flows. We
 compare the results with solutions for homogeneous flows. In the case
 of supersonic flows, the global structure of inhomogeneous flows is not
 much different from that of homogeneous flows. However, detailed
 investigation shows that the average density of inhomogeneous flows
 decreases faster than that of homogeneous flows, because local thermal
 instability removes overdense regions in the inhomogeneous flows and
 reduces the mass flux. We also find that when the gravity of a galaxy
 is strong, the cold clouds formed from the removed gas are distributed
 in the galactic halo. In the case of subsonic flows, the form of
 inhomogeneous flows is different from that of homogeneous flows near
 the regions where the flows terminate. The density rise appearing near
 the regions where the homogeneous flows terminate is not seen in the
 inhomogeneous flows because the local thermal instability decreases the
 mass flux. The cold clouds formed through thermal instability in the
 inhomogeneous flow almost all coast to the same maximum radius.
\end{abstract}


\section{Introduction}
\label{sec:intro}

Some observations suggest that gas clouds in and out of galaxies are
related to outflows from the galaxies.

In the Galactic halo, there are clouds of neutral hydrogen with a
velocity greater than $\sim 80\rm km\;s^{-1}$; they are called high
velocity clouds. There has been an argument about the origin. One idea
is that they result from outflows from the Galaxy; thermal instability
in the flows leads to cloud formation \citep{sha76, bre80, li92}. The
other is that they are made from metal-poor gas falling from the outside
of the Galaxy \citep{oor70, bli99}. Recently, however, \citet{ric99}
found that the iron abundance in a high velocity cloud in the Galactic
halo is half of the solar value, which supports the former idea.

Recent observations of Ly$\alpha$ absorption system show that the
absorbers are associated with galaxies \citep{mor93, lan95, sto95,
bow96, leb96, van96, che98}. For example, the spectroscopy and broadband
imaging of galaxies toward 3C~273 indicate that the Ly$\alpha$ absorbers
are not distributed at random with respected to the galaxies
\citep{mor93}.  \citet{lan95} found an anticorrelation between
Ly$\alpha$ absorption equivalent width and galaxy impact
parameter. Moreover, metals have been found in the absorption system,
which means that the gas is enriched by the products of stellar
nucleosynthesis \citep{rau98}. These observations suggest that the
absorbers are made from the gas ejected from galaxies. In this case,
thermal instability in the outflows is expected to play an important
role in the formation of the absorbers.

\citet{wan95a} investigated cooling gas outflows from galaxies. He
studied the outflow form, the effects of the galaxy potential, the
size of outflow regions, and the efficiency of radiative cooling. He
found that the hot gas in dwarf galaxies can either flow out as galactic
winds, or cool radiatively to form clouds. In the latter case, the
clouds can escape from the galaxies and may become Ly$\alpha$ absorbers
\citep{wan95b}. Recently, \citet{nul98} also indicated that the ejected
gas from dwarf galaxies can give rise to a damped Ly$\alpha$
absorber. In contrast, \citet{wan95a} found that massive galaxies like
our own tend to confine the gas; the gas released into the halo either
cools radiatively or results in a galactic corona.

However, \citet{wan95a} considered only spherical homogeneous
flows. (Here, homogeneous means that density and radius from the galaxy
center have a one-to-one correspondence.) In actual galaxies, however,
there should be density fluctuations that lead to local thermal
instabilities. \citet{bal89} examined the linear evolution of the local
thermal instabilities in outflows and found that the growth is
relatively weak. On the other hand, the non-linear evolution is not
understood well, although complex gas structures observed in galaxies
suggest the fluctuations in outflows are highly non-linear
\citep[e.g.][]{jog98}. Although there are simulations done to study the
non-linear evolution of instability \citep[e.g.][]{kri98}, they cannot
resolve fine structure of gas such as shells of supernova remnants and
gas blobs composed by mass-loss gas from stars; their scales are
typically $\sim 10$ pc and $\sim 1$ pc, respectively \citep{mat90}. If
these fluctuations grow through radiative cooling, the overdensities
condense and are decoupled from outflows. Thus, the substantial
quantities of gas may be removed from the outflows, and the form of
outflows may be changed. As an instance similar to this one, X-ray
observations suggest that a large amount of gas is decoupled from
cooling inflows in the central regions of clusters and the cooled gas is
deposited in the wide region \citep{fab94}. If it happens to be the same
in outflows of a galaxy, the clouds born from the decoupled gas may be
distributed in a wide range of radius of the galaxy.

However, this consideration may be too simplistic; the motion of
overdense regions relative to the ambient gas introduces
complication. If gravity induces the relative motion, the overdense
regions mix may with the ambient gas and cannot cool
anymore. \citet{hat95} numerically showed that the motion strongly
depends on magnetic fields; if the magnetic pressure is strong enough,
it goes against the gravity and the overdensities can be
sustained. \citet{fuj97} estimated that the magnetic pressure could
sustain the overdensities in actual galaxies.

In this paper, we consider the non-linear evolution of fluctuations in
inhomogeneous outflows of galaxies and its influence on the global
structure of outflows. Since it is difficult to treat the non-linear
evolution exactly, we first consider the most extreme case. For this
purpose, we investigate multiphase comoving flows, in which local
thermal instabilities are expected to grow most extremely because the
motion of overdense regions relative to the ambient gas is ignored. We
compare the results with those of homogeneous flows.

Our paper is organized as follows. In \S\ref{sec:model} we summarize our
models for both homogeneous and inhomogeneous flows. In
\S\ref{sec:result} we give the results of our calculations and describe
the differences between homogeneous and inhomogeneous flows. In
\S\ref{sec:disc} the fate of the clouds formed in outflows is
discussed. Conclusions are given in \S\ref{sec:conc}.

\section{Models}
\label{sec:model}
\subsection{Homogeneous Flows}

The equations of mass, momentum, and energy conservation in a steady,
symmetric flow are
\begin{equation}
\frac{1}{r^2}\frac{d(r^2 \rho u)}{dr}=0\:,
\end{equation}
\begin{equation}
u\frac{du}{dr}=-\frac{1}{\rho}\frac{dp}{dr}-\frac{GM}{r^2}\;,
\end{equation}
\begin{equation}
\frac{\rho u}{\gamma-1}\frac{d}{dr}
\ln\left(\frac{p}{\rho^\gamma}\right)=-n_p^2 \Lambda(T)\:,
\end{equation}
where $\rho$, $u$, $p$, are respectively the density, velocity, and
pressure of the gas, $\gamma$ is the adiabatic index, $G$ is the
gravitational constant, $M(r)$ is the gravitating mass within $r$, $n_p$
is the proton density, $\Lambda$ is the cooling function, and $T$ is the
temperature. These equations can be rewritten as dimensionless equations
for the Mach number and the dimensionless temperature (equations [2.14]
and [2.15] in \citealt{wan95a}).

\subsection{Inhomogeneous Flows}

The model adopted here is based on the multiphase cooling flow model of
\citet{nul86}. Contrary to the model of \citet{nul86}, we consider
outflows with relatively large velocity.

Using the volume distribution function $f(\rho, {\bf r}, t)$ at time,
$t$, and position, ${\bf r}$, the mass within the volume $dV$, in the
density range $\rho$ to $\rho+d\rho$ is given by
\begin{equation}
\label{eq:dm}
 dM=\rho f(\rho, {\bf r}, t) d\rho dV.
\end{equation}
Differentiating (\ref{eq:dm}), we obtain the equation of mass
conservation
\begin{equation}
\label{eq:mcons}
 \frac{\partial\rho f}{\partial t}+\nabla\cdot\rho{\bf v}f
+\frac{\partial\rho f \dot{\rho}}{\partial\rho}=0,
\end{equation}
where $\dot{\rho}=d\rho/dt$ is the comoving rate of change of the
density of gas with density $\rho$, and $\bf v$ is the velocity. From
now on, we assume that the flow is steady and spherically symmetric for
simplicity.

The mass flux function $\Psi$ is defined by
\begin{equation}
 \Psi(\rho, r)=\int_0^\rho 4\pi r^2 \rho' u f d\rho'\:,
\end{equation}
where $u$ is the outward radial velocity. As mentioned in
\S\ref{sec:intro}, we assume that the cooling gas comoves, which means
that $u$ is a function of $r$ alone. Thus, there is no local
relative motion between gas blobs before they are decoupled from the
flow. From equation (\ref{eq:mcons}), it can be shown that $\Psi$
satisfies the relation
\begin{equation}
 \dot{\rho}\frac{\partial\Psi}{\partial\rho}
=-u\frac{\partial\Psi}{\partial r}
\end{equation}
\citep[see][]{nul86}. Thus, $\Psi$ is constant along streamlines in the
$\rho, r$ plane.

We assume local pressure equilibrium, that is, pressure, $p$, is also a
function of $r$ alone. In this case, the streamlines for the cooling gas
are determined from two equations; one is the energy equation
\begin{equation}
\label{eq:ene1}
 \rho Tu\frac{dS}{dr}=-n_p^2 \Lambda(T)\:,
\end{equation}
and the other is the momentum equation
\begin{equation}
\label{eq:mom1}
 u\frac{du}{dr}=-\frac{1}{\bar{\rho}}\frac{dp}{dr}
-\frac{GM(r)}{r^2}\:,
\end{equation}
where $S$ is the entropy. The temperature is given by the ratio of $p$
to $\rho$ (or $n_p$). We assume that the cooling function is described
by
\begin{equation}
 \Lambda=D T^\alpha\:,
\end{equation}
where $D$ and $\alpha$ are constants. The spatially averaged density is
given by
\begin{equation}
\label{eq:rho_av}
 \bar{\rho}=\Psi(\infty, r)/\int^\infty_0 
\frac{1}{\rho}\frac{\partial \Psi}{\partial \rho}d\rho \:.
\end{equation}
The total mass flow rate, $\dot{M}$, is given by $\dot{M}=4\pi r^2
\bar{\rho}u$.

We can integrate equation (\ref{eq:ene1}) and give
\begin{equation}
\label{eq:pp}
(p^{0.6}/\rho)^{2-\alpha}=p_{1}^{0.6(2-\alpha)}
(\rho_{i}^{-(2-\alpha)}-\rho_{c}^{-(2-\alpha)})\:,
\end{equation}
where $\rho$ is the density at radius $r$ of the gas with density
$\rho_i$ at radius $r_1$, $p=p(r)$, $p_1=p(r_1)$ and $\rho_c(r)$ is the
density of the gas at $r_1$ which just has infinite density at radius
$r$. The evolution of cooling gas along the streamlines is determined by
$\rho_c$. Using equation (\ref{eq:pp}), equation (\ref{eq:ene1}) is
described by $\rho_c$
\begin{equation}
\label{eq:ene2}
 u\frac{d}{dr}\rho_{c}^{2-\alpha}
=(2-\alpha)\frac{n_{p1}^{2}DT_{1}^{\alpha}}{2.5 p_1}
\rho_{1}^{-(2-\alpha)}\left(\frac{p}{p_1}\right)^{0.2+0.4\alpha}\:,
\end{equation}
where $n_{p1}$ is the number density corresponding to arbitrary density
$\rho_1$, and $T_1$ is the temperature of gas at pressure $p_1$ and
density $\rho_1$.

The energy and momentum equations can be solved by assuming the
functional form of $\Psi$. According to \citet{nul86}, we consider two
cases. One is a power-law form
\begin{equation}
\label{eq:power}
 \Psi(\rho, r_1)=\left\{\begin{array}{ll}
		  A(1-w)^k, & \mbox{$w<1$} \\
		   0,       & \mbox{$w\geq 1$}\:,
		    \end{array}
\right.
\end{equation}
where
\begin{equation}
 w=(\rho/\rho_{\rm m})^{-(2-\alpha)}
\end{equation}
for some constant $\rho_{\rm m}$. Assuming that $k\geq 1$, we obtain
\begin{equation}
\label{eq:rhobar}
 \bar{\rho}=(p/p_1)^{0.6}J\rho_{\rm m}(1-w_c)^{-1/(2-\alpha)}\:,
\end{equation}
where
\begin{equation}
 J=\left(\begin{array}{c}
        k+1/(2-\alpha) \\
	  k
\end{array}
\right)
\end{equation}
and
\begin{equation}
 w_c=(\rho_c/\rho_{\rm m})^{-(2-\alpha)}\:.
\end{equation}

Thus, equation (\ref{eq:mom1}) and (\ref{eq:ene2}) can be rewritten as
\begin{equation}
\label{eq:mom_w}
 \frac{d}{dr}\left(\frac{p}{p_1}\right)=
-\frac{J\rho_{\rm m}}{p_1}
\left(\frac{p}{p_1}\right)^{0.6}(1-w_c)^{-1/(2-\alpha)}
\left(u\frac{du}{dr}+\frac{GM}{r^2}\right)
\end{equation}
and
\begin{equation}
\label{eq:ene_w}
 \frac{d}{dr}(1-w_c)=(2-\alpha)
\frac{4\pi r^2 \rho_{\rm m}J}{At_{\rm cm}}
\left(\frac{p}{p_1}\right)^{0.8+0.4\alpha}(1-w_c)^{-k-1/(2-\alpha)}\:,
\end{equation}
where $t_{\rm cm}$ is the constant pressure cooling time evaluated for
$r=r_1$ and $\rho_1=\rho_{\rm m}$ (or $n_{p1}=n_{\rm pm}$ if represented
by number density), that is,
\begin{equation}
t_{\rm cm}=\frac{5p_1}{2n_{\rm pm}^2 D T_1^{\alpha}}\:.
\end{equation}
Since total mass flow rate is $\dot{M}=A(1-w_c)^k$, equations
(\ref{eq:mom_w}) and (\ref{eq:ene_w}) are respectively given by
\begin{equation}
\label{eq:mom_m}
 \frac{d}{dr}\left(\frac{p}{p_1}\right)=
-\frac{J\rho_{\rm m}}{p_1}
\left(\frac{p}{p_1}\right)^{0.6}
\left(\frac{\dot{M}}{\dot{M}_1}\right)^{-1/k(2-\alpha)}
\left(u\frac{du}{dr}+\frac{GM}{r^2}\right)
\end{equation}
and
\begin{equation}
\label{eq:ene_m}
 \frac{d}{dr}\dot{M}=-k(2-\alpha)
\frac{4\pi r^2 \rho_{\rm m}J}{t_{\rm cm}}
\left(\frac{p}{p_1}\right)^{0.8+0.4\alpha}
\left(\frac{\dot{M}}{\dot{M}_1}\right)^{-1/k-1/k(2-\alpha)}\:, 
\end{equation}
where $\dot{M}_1=\dot{M}(r_1)$. Using equation (\ref{eq:rhobar}), we
obtain 
\begin{equation}
u=\frac{\dot{M}}{4\pi r^2 \bar{\rho}}
 =\frac{\dot{M}}{4\pi r^2 J\rho_{\rm m}}
\left(\frac{p}{p_1}\right)^{-0.6}
\left(\frac{\dot{M}}{\dot{M}_1}\right)^{1/k(2-\alpha)}\:.
\end{equation}
Thus, equations (\ref{eq:mom_m}) and (\ref{eq:ene_m}) are simultaneous
equations for $p$ and $\dot{M}$.

The other form of $\Psi$ is an exponential form, 
\begin{equation}
\label{eq:exp}
\Psi(\rho, r_1)=A e^{-w}\:.
\end{equation}
As is the case of the power-law form (equation [\ref{eq:power}]), we
have the momentum equation
\begin{equation}
\label{eq:p_mom_m}
 \frac{d}{dr}\left(\frac{p}{p_1}\right)=
-\frac{L\rho_{\rm m}}{p_1}
\left(\frac{p}{p_1}\right)^{0.6}
\left(u\frac{du}{dr}+\frac{GM}{r^2}\right)
\end{equation} 
and the energy equation
\begin{equation}
\label{eq:p_ene_m}
 \frac{d}{dr}\dot{M}=-(2-\alpha)
\frac{4\pi r^2 \rho_{\rm m}L}{t_{\rm cm}}
\left(\frac{p}{p_1}\right)^{0.8+0.4\alpha} \:,
\end{equation}
where 
\begin{equation}
L=\frac{1}{(1/(2-\alpha))!} \:.
\end{equation}

\section{Results}
\label{sec:result}

According to \citet{wan95a}, we adopt the cooling function for a gas
with solar metal abundance,
\begin{equation}
\label{eq:coolf}
\Lambda(T)=6.2\times 10^{-19}(T/{\rm K})^{-0.6}\;{\rm cm^{3}\;s^{-1}}
\end{equation}
\citep{mck77, sut93}. This is a good approximation at $10^5\lesssim T
\lesssim 4\times 10^7$~K. Note that this cooling function may not be
appropriate to study the formation of Ly$\alpha$ absorbers because their
observed abundance is about 1/100 solar. However, although the absolute
value of the cooling function with the abundance is about one tenth of
that of equation (\ref{eq:coolf}), the temperature dependence is not
much different from $-0.6$ for $T\lesssim 10^6$~K \citep{sut93}. Thus,
the evolution of the flow with solar abundance is similar to that with
1/100 solar abundance and three or four times larger gas density,
because their local cooling times are almost identical.

For a model galaxy, we consider the isothermal mass distribution
\begin{equation}
\frac{GM}{r}=v_{\rm cir}^2
\end{equation}
with truncated radius $r_{\rm max}$. We consider both supersonic and
subsonic flows. The Mach number of flows, $\cal{M}$, is taken as unity
at the base of the flows. As discussed by \citet{wan95a}, it is a
natural choice at least for the initial condition in supersonic flows,
because a smooth transition from a subsonic flow at the center to a
supersonic flow at large radius requires that $\cal{M}=$1 at $r=R$,
where $R$ is the radius beyond which heating ceases \citep{che85}. The
parameters of model galaxies are the same as those in \citet{wan95a} and
are shown in Table~\ref{tbl-1}. Model D1 corresponds to a typical dwarf
galaxy in which the gas tends to flow out. Model N1 corresponds to a
massive galaxy like our own in which the gas tends to be
confined. Models D2 and N2 are the same as models D1 and N2,
respectively, but the initial gas temperature and the total mass flow
rate are larger. For the power-law type of mass flux function (equation
[\ref{eq:power}]), we take $k=2$ but the results are not much different
even if we take $k=10$.

\subsection{Supersonic Flows}
\label{sec:super}

\begin{figure}
\figurenum{1}
\epsscale{0.80}
\plotone{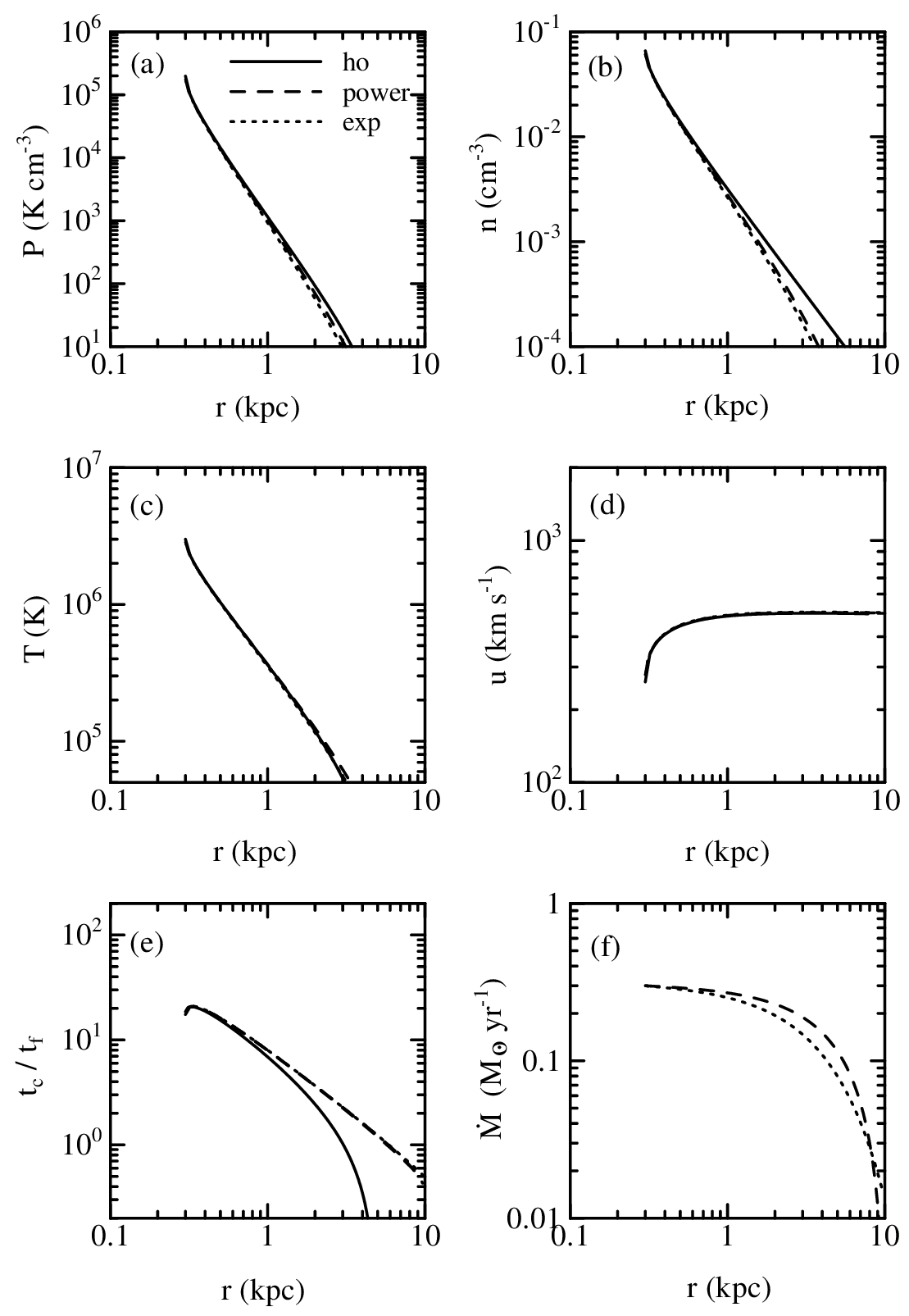}
\caption{(a) Gas pressure, (b) number density, (c)
temperature, (d) velocity, (e) ratio of the cooling time and the flow
time, and (f) total mass flux as functions of radius for supersonic
flows in model~D1. The solid lines indicate homogeneous flows. The
dashed and dotted lines respectively indicate inhomogeneous flows with
the power-law mass-flux function (equation [\ref{eq:power}]; $k=2$) and
those with the exponential mass-flux function (equation
[\ref{eq:exp}]). \label{fig:D1}}
\end{figure}

\begin{figure}
\figurenum{2}
\epsscale{0.80}
\plotone{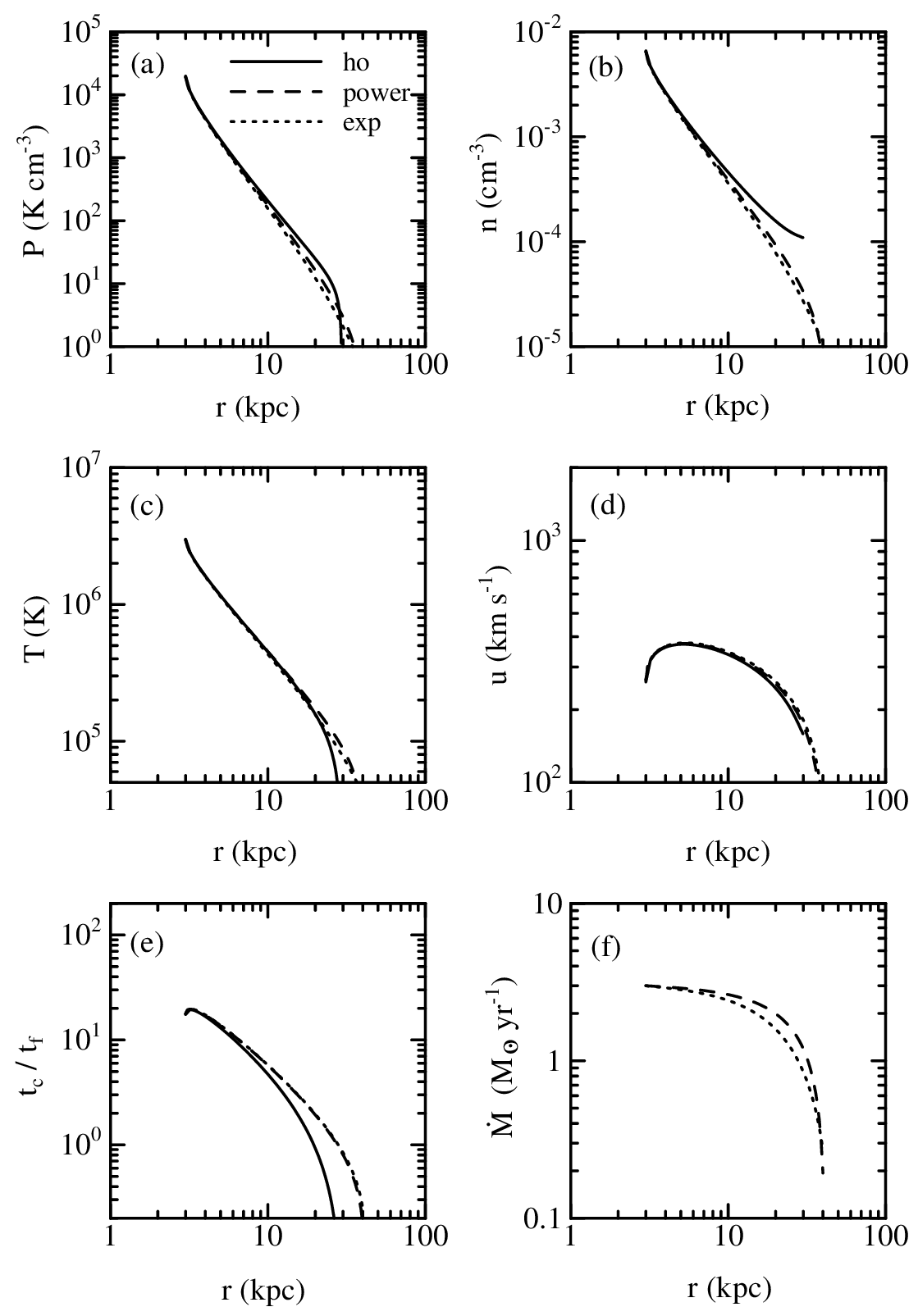}
\caption{The same as Figure~\ref{fig:D1} but for supersonic
flows in model~N1. \label{fig:N1}}
\end{figure}

As an initial condition, we take $\cal M$$=1+\epsilon$ ($\epsilon\sim
10^{-3}-10^{-4}$) at the inner boundary $r_i$. Since the results for
models D2 and N2 are qualitatively similar to those for model D1, we
present only the results for models D1 and N1 here. 
In Figures~\ref{fig:D1} and
\ref{fig:N1}, we show the profiles of pressure, density,
temperature, velocity, and the ratio of cooling to flow time. For
inhomogeneous flows, we plot the average density (equation
[\ref{eq:rho_av}]), and the average temperature given by the ratio of 
the pressure to the average density. The cooling time and flow time 
are defined by
\begin{equation}
\label{eq:cool}
 t_{\rm c}=\frac{P}{(\gamma-1)n_p^2 \Lambda(T)}
\end{equation}
and
\begin{equation}
 t_{\rm f}=\frac{r}{v}
\end{equation}
respectively. In equation (\ref{eq:cool}), we use the averaged density
and temperature for inhomogeneous flows. In Figures~\ref{fig:D1} and
\ref{fig:N1}, we also show the total mass flow rate, $\dot{M}$, for
inhomogeneous flows.

For homogeneous flows, in model~D1 (and D2, N2) hot gas flows out of the
galaxy (Figure~\ref{fig:D1}d), while in model~N1 the hot gas is
confined, although the gas is first accelerated by pressure
(Figure~\ref{fig:N1}d). We define a dimensionless variable
\begin{equation}
 x=\frac{\rho v_{\rm cir}^2}
{4\pi (r/v_{\rm cir}) n_p^2 \Lambda(T)}\:.
\end{equation}
That is, $1/x$ is the fraction of the energy loss due to radiation
during the flow time $r/v_{\rm cir}$ for a flow with velocity equal to
the rotation velocity of the galaxy \citep{wan95a}. At the initial
radius $r_i$, $x\sim 1$ in model N1 and $x<1$ in others. This means that
galactic gravitational potential is important in model N1 while it is
not important in others, which explains the characteristics of the
homogeneous outflows. Since initially $t_{\rm c}>t_{\rm f}$ in model~N1,
the gas first cools adiabatically; radiative cooling becomes dominant
later on ($t_{\rm c}< t_{\rm f}$). Note that in model~D1 although
$t_{\rm c}/t_{\rm f}$ also becomes smaller than one for $r\gtrsim 3$~kpc
for homogeneous flow (Figure~\ref{fig:D1}e), the gas temperature of the
flow has already been smaller than $10^5$~K at $r\sim 2$~kpc
(Figure~\ref{fig:D1}c), that is, equation (\ref{eq:coolf}) should not be
adopted there.

Broadly speaking, the profiles of pressure, average density, average
temperature, and velocity of the hot gas for inhomogeneous flows are
similar to those for homogeneous flows. However, it is shown that the
gas pressure and density for the inhomogeneous flows decrease rather
faster than those for the homogeneous flows (Figures~\ref{fig:D1} and
\ref{fig:N1}). This is because of the decrease of the mass flux
(equations [\ref{eq:ene_m}] and [\ref{eq:p_ene_m}]). In fact,
Figures~\ref{fig:D1}f and \ref{fig:N1}f show that $\dot{M}$ decreases
outwardly, reflecting that thermal instability removes overdense regions
from the flows.  In model~D1, in the region where $\dot{M}$ decreases
($r\gtrsim 1$~kpc), the terms of pressure gradient and gravity in
equation (\ref{eq:mom1}) are small and the gas expands freely ($u\sim
constant$). Thus, the decrease of $\dot{M}$ reduces $\bar{\rho}$ because
$\bar{\rho}=\dot{M}/(4\pi r^2 u)$. In model~N1, in the region where
$\dot{M}$ decreases ($r\gtrsim 7$~kpc), the gravity term in equation
(\ref{eq:mom1}) dominates the pressure gradient term. This means that
the velocity $u$ is determined only by the gravitational field and does
not depend on the homogeneity of the flow. Thus, the decrease of
$\dot{M}$ reduces $\bar{\rho}$ as is the case of model~D1. It is to be
noted that even in the region where $t_{\rm c}/t_{\rm f}>1$, $\dot{M}$
decreases (Figures~\ref{fig:D1} and \ref{fig:N1}), because $t_{\rm c}$
is defined for the averaged density and the overdense regions in the
flows cool ahead of most of the gas.

\subsection{Subsonic Flows}

The initial conditions for subsonic flows are the same as those of the
supersonic flows except for the initial velocity; we take $\cal
M$$=1-\epsilon$ ($\epsilon\sim 10^{-3}-10^{-4}$) at the inner boundary
$r_i$. We present the results for models~D1 and N1; the results for
other models are qualitatively the same as those for model~D1.

\begin{figure}
\figurenum{3}
\epsscale{0.80}
\plotone{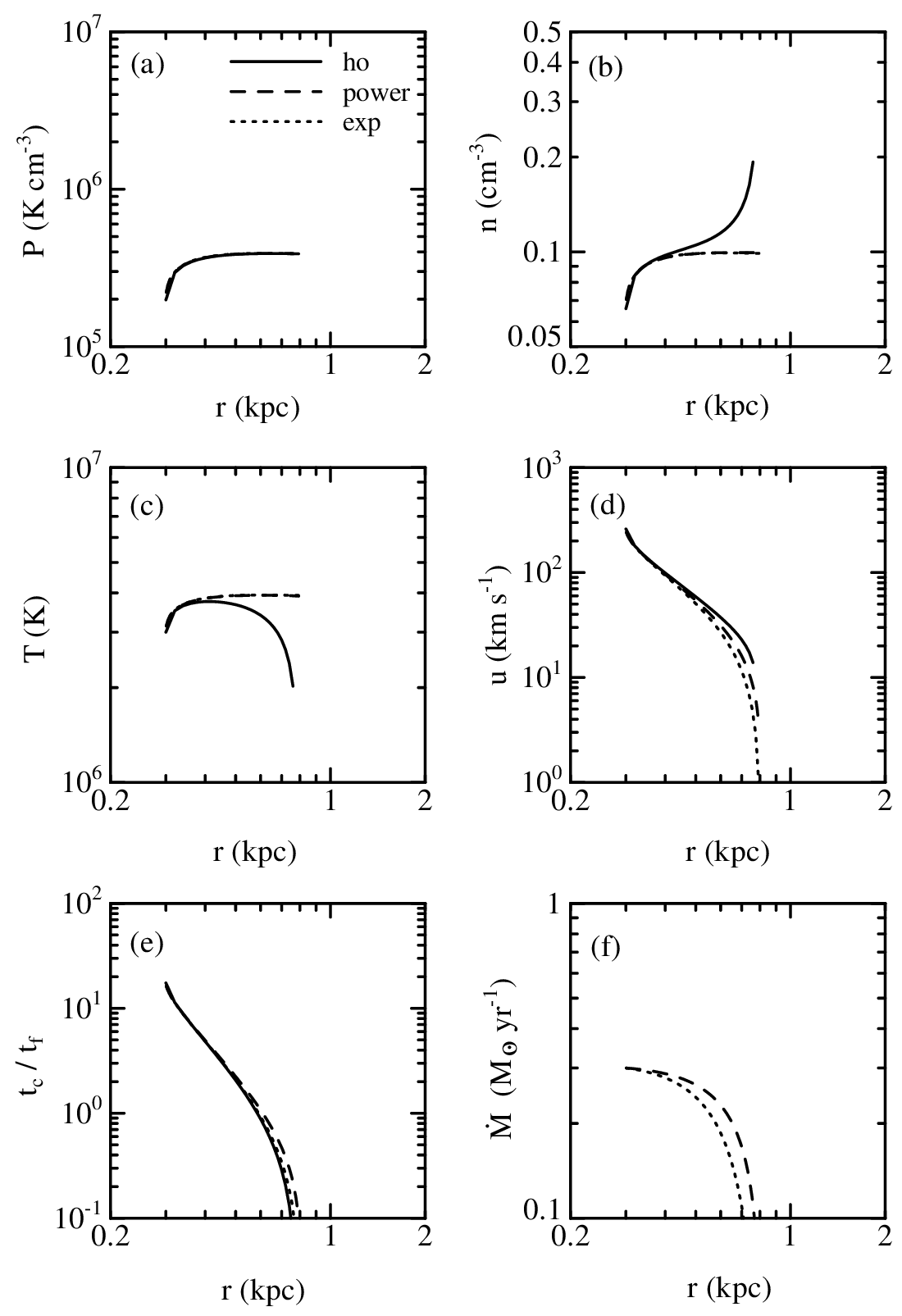}
\caption{The same as Figure~\ref{fig:D1} but for subsonic
flows in model~D1. \label{fig:sD1}}
\end{figure}

\begin{figure}
\figurenum{4}
\epsscale{0.80}
\plotone{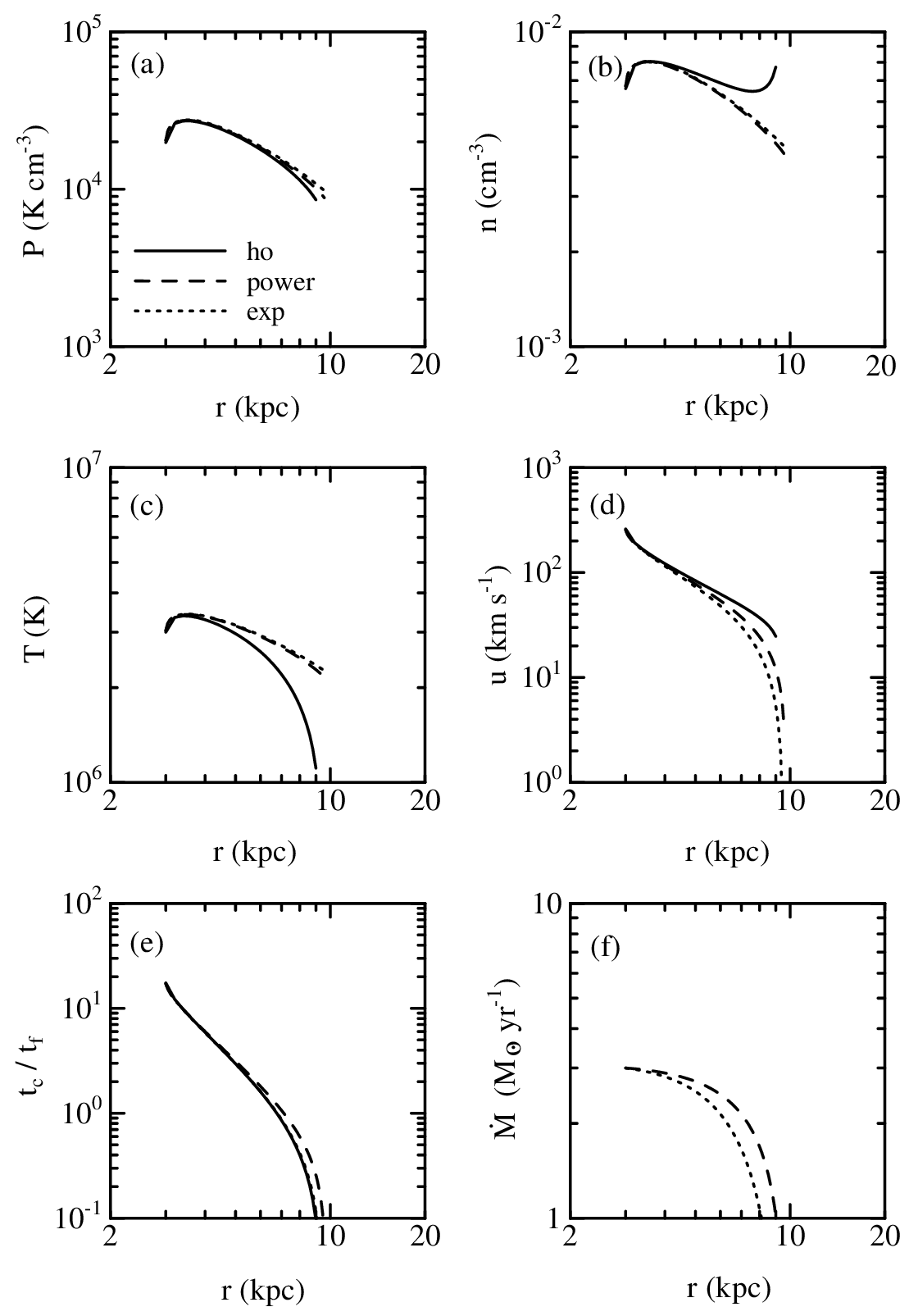}
\caption{The same as Figure~\ref{fig:D1} but for subsonic
flows in model~N1. \label{fig:sN1}}
\end{figure}

Although gas is confined in a galaxy for all models, the solutions for
inhomogeneous flows are different from those for homogeneous flows
(Figures~\ref{fig:sD1} and \ref{fig:sN1} ). The density increase
appearing near the regions where the homogeneous flows terminate does
not exist for the inhomogeneous flows. Figures~\ref{fig:sD1} and
\ref{fig:sN1} show that $t_{\rm c}/t_{\rm f}$ decreases rapidly because
the velocity of the flows reduces. Thus, radiative cooling becomes
important and causes gas condensation near the regions where the
homogeneous flows terminate. On the other hand, for inhomogeneous flows,
radiative cooling makes mass-dropout effective and prevents the density
increase from appearing.

In model~N1 in \S\ref{sec:super} and in the models discussed in this
subsection, radiatively cooled gas piles up at finite distances from a
galaxy. This appears to be inconsistent with the assumption of steady
flows. However, if the cold gas turns into dense, compact clouds, the
clouds ballistically fall back to the galaxy and the gas accumulation
does not significantly affect the global structure of the outflows
unless they are disrupted and mix with ambient gas soon after the
formation. The stability of the clouds have been studied by many
authors. For example, although the clouds are subject to
Kelvin-Helmholtz instability, \citet{mur93} indicated that the clouds are
stable for a long time if they are bound by a sufficiently strong
(self-)gravitational potential. Moreover, \citet{hat00} indicated that
the evaporation of the clouds by surrounding hot gas may be greatly
suppressed by a plasma instability. In the next section, we discuss the
motion of the clouds including the case that the gravity of a galaxy
cannot stop outflows.

\section{Discussion}
\label{sec:disc}

We assume that cooled gas in outflows turns to be stable clouds, which
move ballistically. The initial velocity of the newly formed clouds is
expected to be the velocity of the flow at the position of their
formation if the formation time-scale is small enough. We consider
whether the clouds escape from a galaxy or fall back under the influence
of the galactic potential.

Since the mass distribution of our model galaxy is truncated at radius
$r_{\rm max}$, the escape velocity is given by
\begin{equation}
v_{\rm esc}(r)=\sqrt{2}\:v_{\rm cir}
\left[1+\ln\left(\frac{r_{\rm max}}{r}\right)\right]
\end{equation}
for $r<r_{\rm max}$. We take $r_{\rm max}=50\: r_{\rm i}$ from now
on. If the clouds cannot escape from the galaxy, the maximum radius they
can reach is given by
\begin{equation}
\label{eq:rf1}
r_{\rm f}=r_{\rm c}\exp\left[\frac{1}{2}
\left(\frac{v_{\rm c}}{v_{\rm cir}}\right)^2\right]\:,
\end{equation}
if $r_{\rm f}\leq r_{\rm max}$, or
\begin{equation}
\label{eq:rf2}
r_{\rm f}=\frac{2 v_{\rm cir}^2}
{v_{\rm esc}^2(r_{\rm c})-v_{\rm c}^2}r_{\rm max}\:,
\end{equation}
if $r_{\rm f}> r_{\rm max}$, where $v_{\rm c}$ is the velocity of
clouds when they form at radius $r_{\rm c}$ \citep{wan95a}. In equations
(\ref{eq:rf1}) and (\ref{eq:rf2}), we assume that $r_{\rm c}<r_{\rm
max}$. For all the models adopted here, this relation is satisfied.

\begin{figure}
\figurenum{5}
\epsscale{0.80}
\plotone{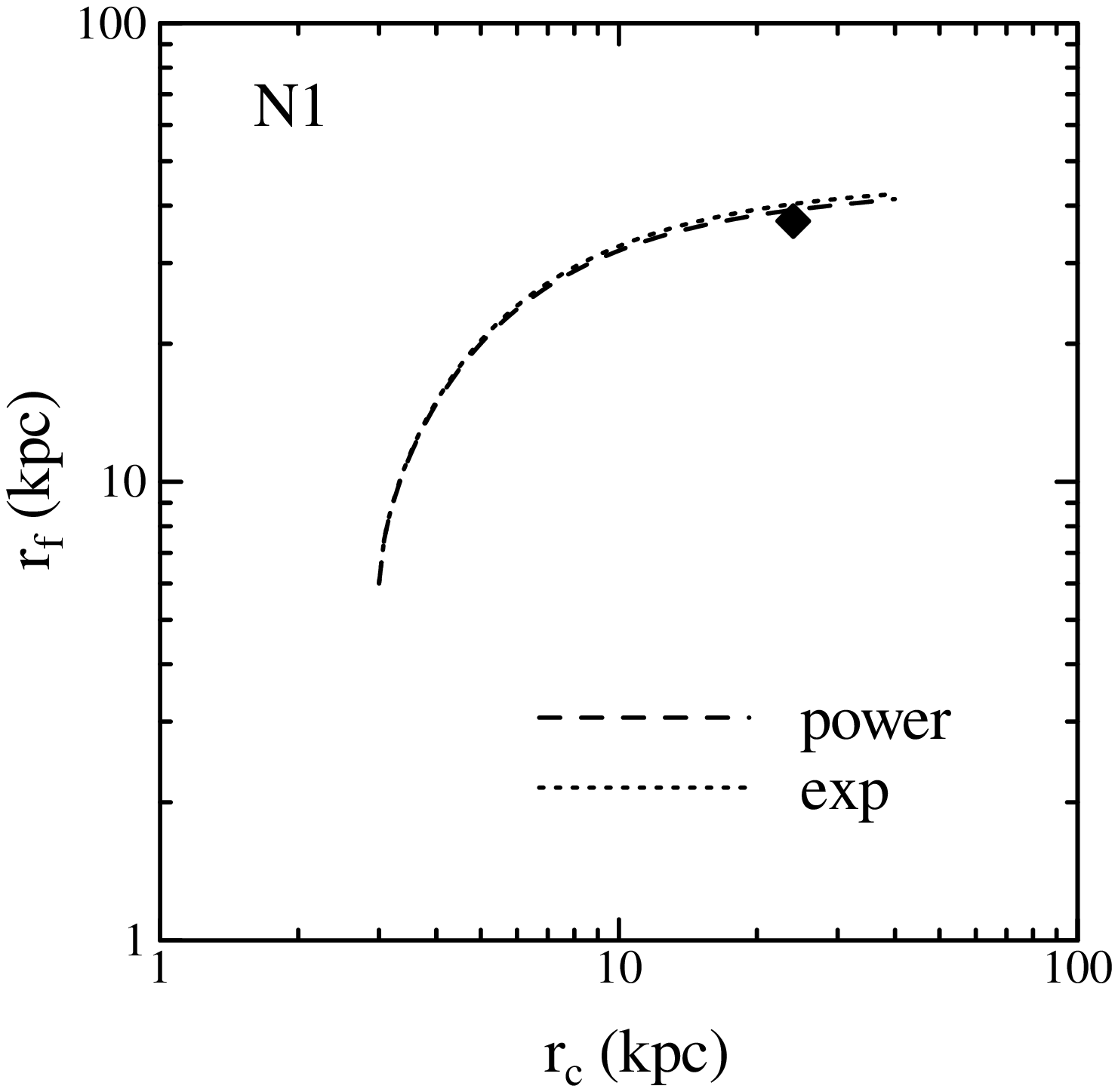}
\caption{The relation between the position where clouds form
and the position where the clouds reach for supersonic flows in
model~N1. The dashed and dotted lines respectively indicate
inhomogeneous flows with the power-law mass-flux function (equation
[\ref{eq:power}]; $k=2$) and those with the exponential mass-flux
function (equation [\ref{eq:exp}]). The diamond shows the result for the
homogeneous flow. \label{fig:sup}}
\end{figure}

\begin{figure}
\figurenum{6}
\epsscale{0.80}
\plotone{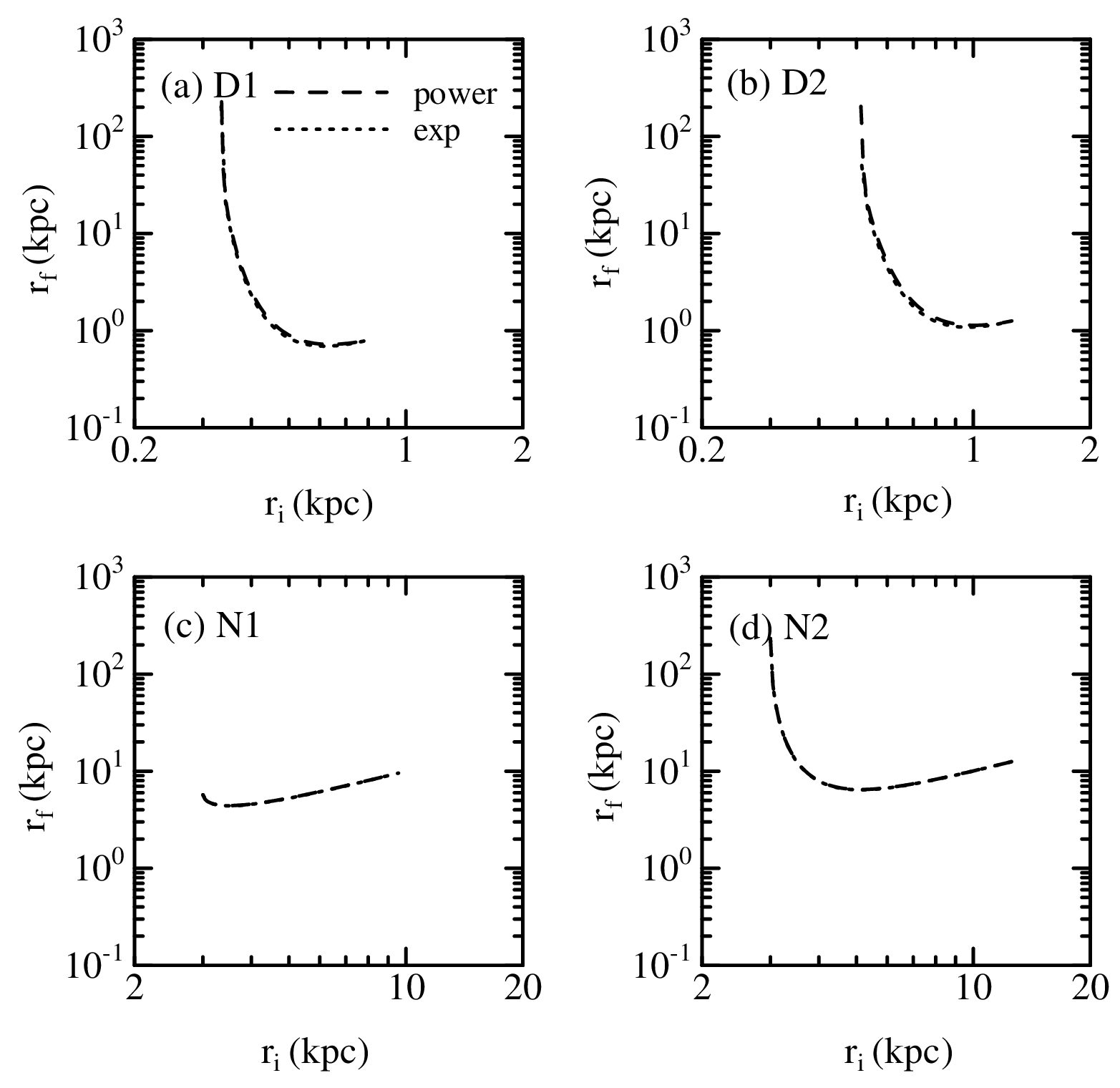}
\caption{The relation between the position where clouds form
and the position where the clouds reach for supersonic flows. The dashed
and dotted lines respectively indicate inhomogeneous flows with the
power-law mass-flux function (equation [\ref{eq:power}]; $k=2$) and
those with the exponential mass-flux function (equation
[\ref{eq:exp}]). (a) model~D1 (b) model~D2 (c) model~N1 (d)
model~N2. \label{fig:sub}}
\end{figure}

Figures~\ref{fig:sup} and \ref{fig:sub} show the relation between
$r_{\rm c}$ and $r_{\rm f}$ for supersonic and subsonic flows,
respectively. First, we discuss supersonic flows. Since in models D1,
D2, and N2, the clouds can escape from the galaxy regardless of the
position where they are born, we discuss only model N1. For the
homogeneous flow, assuming that clouds form when the temperature of gas
drops to $\sim 10^5$~K, the clouds cannot escape form the galaxy
(Figure~\ref{fig:sup}). For the inhomogeneous flows, the clouds also
cannot escape but $r_{\rm f}$ has a distribution. The clouds that form
nearer to the galaxy center have smaller $r_{\rm f}$. By comparing
Figure~\ref{fig:sup} with Figure~\ref{fig:N1}, it is shown that about
two third of the clouds formed from the inhomogeneous flows have the
orbit which is almost the same as the clouds formed from the homogeneous
flow ($r_{\rm f}\sim 40$~kpc); other clouds have smaller apocentric
radii.

Second, we consider subsonic flows. For homogeneous flows, the initial
velocity of clouds is almost zero, if we assume that they form when the
temperature of gas drops to $\sim 10^5$~K. Thus, the clouds start to
fall back to the galaxy immediately. For inhomogeneous flows, most of
the clouds cannot escape from the galaxy, although the clouds that form
near to $r_{\rm i}$ can escape in models~D1 and D2
(Figure~\ref{fig:sub}). Contrary to the supersonic flows in model~N1
(Figure~\ref{fig:sup}), the clouds that form nearer to the galaxy have
larger $r_{\rm f}$ in models~D1, D2, and N2, because $v_{\rm c}$ is a
strong decreasing function of $r$. However, in model~D1 for example, in
spite of the wide distribution of $r_{\rm f}$, it can be shown that more
than 90\% of the gas which originally composed of the outflow can coast
only to $r_{\rm f}\sim 1$~kpc by comparing Figure~\ref{fig:sub}a with
Figure~\ref{fig:sD1}f. This can be said to other models as well; most of
the gas dropped out of the flow reaches the almost same radius ($r_{\rm
f}\sim 1$~kpc for model~D2 and $r_{\rm f}\sim 10$~kpc for models~N1 and
N2). Moreover, the subsonic flows cannot explain the observed
correlation between galaxies and Ly$\alpha$ absorbers because their
separations range 12.4 to 157.4$h^{-1}$ \citep{che98}.

The clouds formed in the galactic halo fall back toward the galaxy. For
our model galaxy, a cloud with velocity zero at radius $r_{\rm f}$ has
infall velocity,
\begin{equation}
 v_{\rm fall}=v_{\rm cir}
\sqrt{2\ln\left(\frac{r_{\rm f}}{r_{\rm fall}}\right)} \:,
\end{equation}
at radius $r_{\rm fall}$ if $r_{\rm f}<r_{\rm max}$. When $v_{\rm
cir}=225\;\rm km\; s^{-1}$ and $r_{\rm f}=10$~kpc, $v_{\rm
fall}=128\;\rm km\; s^{-1}$ at $r_{\rm fall}=8.5$~kpc. Thus, the
observed high velocity clouds may be related to the clouds formed from
the subsonic flows.

\section{Conclusions}
\label{sec:conc}

We study thermal instability in galactic outflows (or winds) and its
effect on the global structure of the outflows. In order to investigate
the maximum effect, we consider comoving inhomogeneous (multiphase)
outflows in which overdense regions move together with ambient gas;
comoving flows may be possible if magnetic fields are strong enough.

We formulate the inhomogeneous flow and apply it to dwarf galaxies and
normal galaxies. We compared the results with solutions for homogeneous
flows. For supersonic flows, the pressure, density, temperature, and
velocity profiles of hot gas for homogeneous and those for inhomogeneous
flows are similar. However, strictly speaking, the density of
inhomogeneous flows decreases somewhat faster than that of homogeneous
flows as the gas expands. This is because local thermal instability
removes overdensities from the flows owing to local gas condensation. On
the other hand, for subsonic flows, solutions for inhomogeneous flows
are significantly different from those of homogeneous flows near the
regions where the flows terminate. While radiative cooling and gas
condensation raise gas density near the regions where the homogeneous
flows terminate, local thermal instability in the inhomogeneous flows
removes overdensities and prevents an upturn of the average density.

The gas radiatively cooled is most likely to form clouds. The newly
formed clouds inherit the velocity of a flow at the position of their
formation if the time-scale of the formation is much smaller than the
flow time-scale. For supersonic flows, we find that the clouds can
escape from dwarf galaxies regardless of the homogeneity of flows. On
the other hand, for massive galaxies like our own, the clouds cannot
escape from the galaxies without an extreme starburst; in particular,
the model of an inhomogeneous supersonic flow predicts that the clouds
are widely distributed in a galactic halo. For subsonic flows, most of
the clouds are confined in galaxies regardless of the homogeneity of
flows and the galaxy mass. Most of the clouds formed in an inhomogeneous
flow reach the almost same radius before being pulled back by
gravitation.

Although the assumption of comoving flows may be too extreme and the
effect of thermal instability estimated here may be too strong, we think
that this possibility cannot be ruled out at present.

\acknowledgments

I am grateful to K. Okoshi and S. Inutsuka for useful
discussions. Comments from an anonymous referee led to significant
improvements in the quality of this paper.

\clearpage

\begin{deluxetable}{ccccc}
\footnotesize
\tablecaption{Model Parameters. \label{tbl-1}}
\tablewidth{0pt}
\tablehead{
\colhead{Models} & \colhead{$v_{\rm cir}$} 
& \colhead{$r_i$}   & \colhead{$T_i$}   &
\colhead{$\dot{M}(r_i)$} \\
\colhead{}  & \colhead{$(\rm km\;s^{-1}$)} & 
\colhead{(kpc)}     & \colhead{(K)}  &
\colhead{($\rm M_{\sun}\; yr^{-1}$)}   
}
\startdata
D1 &50 &0.3 &$3\times 10^6$& 0.3 \\
D2 &50 &0.3 &$2\times 10^7$& 10  \\
N1 &225&3   &$3\times 10^6$& 3   \\
N2 &225&3   &$2\times 10^7$& 100 \\
\enddata

\end{deluxetable}

\clearpage

\end{document}